\documentclass[aps,prl,reprint,superscriptaddress,showpacs,preprintnumbers,amsmath,amssymb]{revtex4-1}
\usepackage{graphicx}
\usepackage{dcolumn}
\usepackage{bm}
\usepackage{color}
\begin{document}
\preprint{K. Saitoh and B.P. Tighe}
\title{Non-local effects in inhomogeneous flows of soft athermal disks}
\author{Kuniyasu Saitoh}
\email[]{kuniyasu.saitoh.c6@tohoku.ac.jp}
\affiliation{Research Alliance Center for Mathematical Sciences, Tohoku University, 2-1-1 Katahira, Aoba-ku, Sendai 980-8577, Japan}
\affiliation{WPI-Advanced Institute for Materials Research, Tohoku University, 2-1-1 Katahira, Aoba-ku, Sendai 980-8577, Japan}
\author{Brian P. Tighe}
\affiliation{Delft University of Technology, Process \& Energy Laboratory, Leeghwaterstraat 39, 2628 CB Delft, The Netherlands}
\date{\today}
\begin{abstract}
We numerically investigate non-local effects on inhomogeneous flows of soft athermal disks close to but below their jamming transition.
We employ molecular dynamics to simulate Kolmogorov flows, in which a sinusoidal flow profile with fixed wave number is externally imposed, resulting in a spatially inhomogeneous shear rate.
We find that the resulting  rheology is strongly wave number-dependent, and that particle migration, while present, is not sufficient to describe the resulting stress profiles within a conventional local model.
We show that, instead, stress profiles can be captured with non-local constitutive relations that account for gradients to fourth order.
Unlike nonlocal flow in yield stress fluids, we find no evidence of a diverging length scale.
\end{abstract}
\maketitle
%
Predictive descriptions of the rheology of soft athermal particles,\ e.g.\ emulsions, foams, colloidal suspensions, and granular materials,
are frequently needed in the context of food, pharmaceutical, personal care products, and other process technologies \cite{review-rheol0}.
Recently, physicists have studied the constitutive relations of these out-of-equilibrium systems in the context of jamming or yielding transitions \cite{rheol0,rheol7,pdf1,rheol2,rheol3,rheol10,saitoh11,baumgartenSM2017,clarkPRE2018}.
However, nearly all effort to-date has addressed homogeneously flowing systems, and the resulting local constitutive relations \cite{muI0},
even if they are generalized to tensorial forms \cite{muI2}, are blind to so-called \emph{non-local effects} \cite{Eringen}
which are relevant to spatially inhomogeneous flows of disordered materials \cite{review-rheol1,nonlocal0}.

Phenomenologically, non-locality in flow refers to constitutive relations that are sensitive to spatial gradients in the shear rate. 
In dense amorphous matter, the effect is presumed to result from plastic events triggered by distant stress fluctuations \cite{nonlocal1,nonlocal2,nonl-pl0,nonl-pl2,nonl-pl3,nonl0}.
In recent years there has been substantial interest in the nonlocal continuum model of Bocquet and co-workers \cite{sus-nonl0} and several related models \cite{nonlocal1,nonl1,nonl2,nonl3,nonl4,bouzid15}. They take the usual local constitutive relation, determined under homogeneous flow conditions, and introduce it as a source term in a diffusion equation for the fluidity (inverse viscosity). A so-called ``cooperativity length'' is required to quantify the range of non-local effects. These models successfully describe inhomogeneous flow profiles in emulsions \cite{sus-nonl0,sus-nonl1,sus-nonl2}, foams \cite{foam-nonl0}, and granular materials \cite{nonl1,nonl2,nonl4,tang18} under conditions where local models fail dramatically. 

Despite these successes, important questions remain regarding how and when non-local effects are significant. The original fluidity model incorporated a cooperativity length that vanishes as the volume fraction $\phi$ approaches the jamming volume fraction $\phi_J$ from above \cite{sus-nonl0}.
In sharp contrast, more recent efforts call for a length scale that diverges at a critical stress \cite{nonlocal1,nonl1,nonl2,nonl3,nonl4,bouzid15}.
Though mutually inconsistent, both approaches predict that non-locality requires a yield stress.
Cagny et al.~probed granular suspensions without a yield stress and found that velocity profiles can also be fit with the fluidity model, albeit with a cooperativity length proportional to the rheometer's gap width \cite{con-coupling1}.
They argued the length scale is merely a proxy for particle migration effects, and showed that a local model can describe the profiles if one accounts for spatial variations in the viscosity.
Hence the applicability of nonlocal models {\em below} jamming remains uncertain.

In this Letter, we study non-local effects in \emph{Kolmogorov flow}, in which the system flows steadily under forcing that varies sinusoidally in space.
This method builds on prior work in liquids \cite{todd08}, granular materials \cite{schulz03,kuhn05}, and foams and emulsions \cite{nonl-el0}.
We simulate dense systems of soft, viscous, athermal disks \cite{fo1}, the canonical model of jamming.
Prior studies of this system have focused on homogeneous flows, and have evidenced a sensitive (critical) dependence of the homogeneous flow curves on both the proximity to jamming, $\Delta\phi=\phi_J-\phi$, and the shear rate, $\dot{\gamma}$ \cite{rheol0,rheol7,pdf1}.
From our own simulations of simple shear flows, we have verified that both the shear stress and normal stress,\
i.e.\ $\sigma_{xy}^\mathrm{L}=\eta_s(\phi,\dot{\gamma})\dot{\gamma}$ and $\sigma_{yy}^\mathrm{L}=\eta_c(\phi,\dot{\gamma})\dot{\gamma}$,
can be described with the viscosity,
\begin{equation}
\eta_o(\phi,\dot{\gamma}) =
\begin{cases}
\bar{\eta}_o(\dot{\gamma}^{a_o}+c_o\Delta\phi^{b_o})^{-1} & (\phi<\phi_J) \\
\sigma_o(\phi)\dot{\gamma}^{-1}+\bar{\eta}_o\dot{\gamma}^{-a_o} & (\phi>\phi_J)
\end{cases}
\label{eq:constitutive}
\end{equation}
($o=s,c$), where we summarize the yield stress, $\sigma_o(\phi)$, and fitting parameters, $\bar{\eta}_o$, $a_o$, $b_o$, and $c_o$, in Supplemental Materials (SM) \cite{SupplMater}.
Our focus here is primarily on the case without a yield stress, $\phi < \phi_J \simeq 0.842$.
We find (i) constitutive relations depend on gradients of the strain rate; (ii) particle migration modifies the predictions of local models, but cannot account for the observed stress profiles; (iii) non-local models correctly capture the resulting stress profiles; while (iv) the cooperativity length remains small for all simulated flow parameters.

\begin{figure}
\includegraphics[width=\columnwidth]{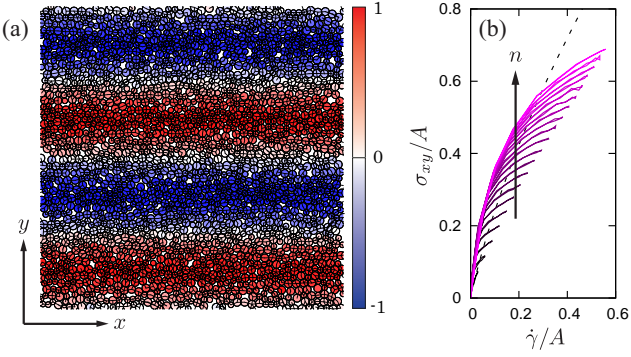}
\caption{(Color online)
(a) Snapshot of  Kolmogorov flow with wave number $n=2$. Colors represent the velocity, $-1\le v_{ix}/A\le 1$.
Solid lines have width proportional to the  elastic forces between the $N=2048$ disks (circles).
(b) Flow curves obtained from Kolmogorov flow at $\phi=0.82$. Lines represent wave numbers $q_n = 2\pi n/L$ increasing from $n=1$ to $20$.
Shear stress, $\sigma_{xy}$, and shear rate, $\dot{\gamma}$, are scaled by the amplitude, $A=10^{-3}d_0/t_0$, and the dotted line represents \ Eq.\ (\ref{eq:constitutive}).
\label{fig:setup}}
\end{figure}
\emph{Numerical methods.}---
We use MD simulations of two-dimensional disks.
To generate initial disordered configurations, we randomly distribute a 50:50 binary mixture of $N=131072$ disks in a $L\times L$ square periodic box.
Here, different kinds of disks have different diameters, $d_L$ and $d_S$, with their ratio, $d_L/d_S=1.4$, so that area fraction is given by $\phi\equiv\pi(d_L^2+d_S^2)N/8L^2$.
Repulsive forces between contacting disks are modeled by linear elastic forces,\
i.e.\ $\mathbf{f}_{ij}^\mathrm{el}=k(R_i+R_j-r_{ij})\mathbf{n}_{ij}$ for $R_i+R_j>r_{ij}$ and $\mathbf{f}_{ij}^\mathrm{el}=\mathbf{0}$ otherwise,
where $R_i$ labels the radius of disk $i$ and $r_{ij}$ is the center-to-center distance between the disks $i$ and $j$.
In the elastic force, $k$ and $\mathbf{n}_{ij}\equiv\mathbf{r}_{ij}/r_{ij}$ with the relative position, $\mathbf{r}_{ij}\equiv\mathbf{r}_i-\mathbf{r}_j$, represent the stiffness and normal unit vector, respectively.
The system is relaxed to a static state by means of FIRE algorithm \cite{FIRE}.
In order to simulate flow, we add viscous forces to every disk as $\mathbf{f}_i^\mathrm{vis}=-\eta\left\{\mathbf{v}_i-\mathbf{u}(\mathbf{r}_i)\right\}$,
where $\eta$, $\mathbf{v}_i$, and $\mathbf{u}(\mathbf{r})$ are the bulk viscosity, velocity of disk $i$, and external flow field, respectively.
Then, we describe motions of the disks by overdamped dynamics \cite{rheol0,rheol7,pdf1},\ i.e.\ $0 = \sum_{j\neq i}\mathbf{f}_{ij}^\mathrm{el}+\mathbf{f}_i^\mathrm{vis}$,
such that the velocity is given by $\mathbf{v}_i = \mathbf{u}(\mathbf{r}_i)+\eta^{-1}\sum_{j\neq i}\mathbf{f}_{ij}^\mathrm{el}$.
In the following analyses, we scale every length and time by the units, $d_0\equiv(d_L+d_S)/2$ and $t_0\equiv\eta/k$, respectively, and change the area fraction from $\phi=0.80$ to $0.85$.

To simulate Kolmogorov flow, we apply flow fields $\mathbf{u}(\mathbf{r})=(u_n(y),0)$, where the $x$-component is given by 
\begin{equation}
u_n(y) = A\sin q_n y~,
\label{eq:kolm}
\end{equation}
with an amplitude, $A$, and wave number, $q_n\equiv 2\pi n/L$ ($n=1,2,\dots$).
The flow field,\ Eq.\ (\ref{eq:kolm}), is periodic along the $y$-axis, and we use periodic boundary conditions to avoid non-local effects due to boundaries \cite{bound-nonl0}.
We take time-averages over the interval $20\le At/d_0\le 50$, which we have verified to be in steady state \cite{SupplMater}.

\begin{figure}
\includegraphics[width=\columnwidth]{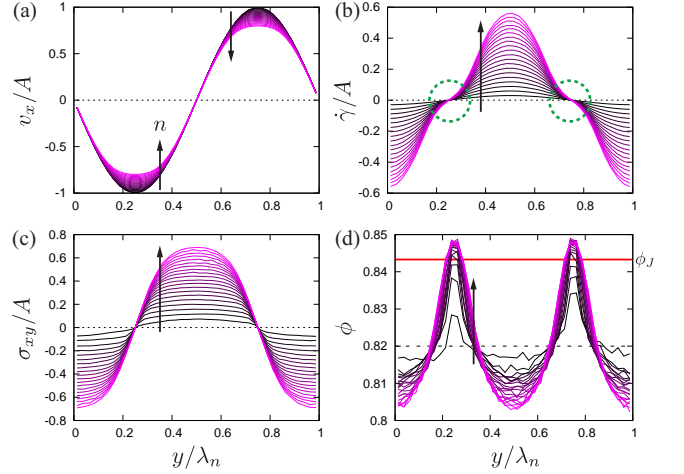}
\caption{(Color online)
Profiles of (a) the velocity field, $v_x(y)/A$, (b) shear rate, $\dot{\gamma}(y)/A$, (c) shear stress, $\sigma_{xy}(y)/A$, and (d) area fraction, $\phi(y)$,
where $A$, $\phi$, and $n$ are as in Fig.\ \ref{fig:setup}(b).
Wave numbers increase in the direction of the arrows.
Dotted circles in (b) indicate shear localized regions and the red solid line in (d) represents the jamming point, $\phi_J$.
\label{fig:profiles}}
\end{figure}
\emph{Breakdown of local rheology.}---
First, we examine the local rheology of Kolmogorov flows.
Figure \ref{fig:setup}(a) shows a steady state flow with $n=2$.
In this figure, force-chains (the solid lines) develop around nodes of the sinusoidal flow field,\ Eq.\ (\ref{eq:kolm}), so that the elastic forces, $\mathbf{f}_{ij}^\mathrm{el}$, do not vanish
and velocities of the disks, $\mathbf{v}_i = \mathbf{u}(\mathbf{r}_i)+\eta^{-1}\sum_{j\neq i}\mathbf{f}_{ij}^\mathrm{el}$, can deviate from the flow field, $\mathbf{u}(\mathbf{r})$.
This means that the local shear rate is different from $\nabla_y u_n(y)=Aq_n\cos q_n y$ and the stress will show non-trivial local profiles
(in contrast with previous studies \cite{sus-nonl0,sus-nonl1,sus-nonl2,nonl0,nonl1,nonl2,nonl4}, where the stress profiles are statically determinate).
We compute the velocity field by dividing the system into small bins as $v_x(y)=N(y)^{-1}\sum_{y_i\in y}v_{ix}$,
where the summation runs over the $N(y)$ disks located in a bin,\ i.e.\ the disks satisfying $y_i\in[y-\Delta y/2,y+\Delta y/2]$ with the bin size, $\Delta y$.
We then take the $y$-derivative of the velocity field such that the local shear rate is given by $\dot{\gamma}(y)=\nabla_y v_x(y)$.
Similarly, the shear stress is calculated according to $\sigma_{xy}(y)=(L\Delta y)^{-1}\sum_{y_i\in y}\sum_{j\neq i}f_{ijx}^\mathrm{el}y_{ij}$
which corresponds to the macroscopic shear stress if we integrate it over the entire system,\ i.e.\ $\sigma_{xy}^\mathrm{L}=L^{-1}\int_{-L/2}^{L/2}\sigma_{xy}(y)dy$ \cite{SupplMater}.

If the constitutive relations, Eq.\ (\ref{eq:constitutive}), are applicable to inhomogeneous flows,
the shear stress, $\sigma_{xy}(y)$, must respond to the local shear rate, $\dot{\gamma}(y)$, in the same way as the macroscopic shear stress, $\sigma_{xy}^\mathrm{L}$.
Figure \ref{fig:setup}(b) shows parametric plots of $\sigma_{xy}(y)$ and $\dot{\gamma}(y)$ for $\phi=0.82$ and $A=10^{-3}d_0/t_0$.
In this figure, we increase the wave number from $n=1$ to $20$ (arrow), where the dotted line is the response of macroscopic shear stress, $\sigma_{xy}^\mathrm{L}$, which we consider as the limit of $n=0$.
Clearly, variations of the flow fields are significant and the flow curves are wave-number dependent.
Therefore, the local constitutive relations, Eq.\ (\ref{eq:constitutive}), fail to describe Kolmogorov flow.

\emph{Particle migration.}---
What is the origin of the wave-number dependence seen in Fig.~\ref{fig:setup}?
As suggested by de Cagny et al. \cite{con-coupling1}, we now examine the role of particle migration.
Figure \ref{fig:profiles} displays the profiles of (a) velocity field, $v_x(y)$, (b) shear rate, $\dot{\gamma}(y)$, and (c) shear stress, $\sigma_{xy}(y)$,
where the amplitude, $A$, area fraction, $\phi$, and wave numbers, $n$, are as in Fig.\ \ref{fig:setup}(b), and the $y$-coordinate is scaled by the wavelength, $\lambda_n\equiv L/n$.
Increasing the wave number from $n=1$ to $20$ (as indicated by the arrows), we find that the velocity field around anti-nodes is flattened (Fig.\ \ref{fig:profiles}(a))
and accordingly the shear rate becomes small (in the dotted circles in Fig.\ \ref{fig:profiles}(b)),\ i.e.\ Kolmogorov flow with high wave numbers exhibits shear localization.
Fig.\ \ref{fig:profiles}(d) shows area fraction profiles, $\phi(y)$, which vary significantly in the vicinity of shear localization.
Hence particle migration is indeed present.

\begin{figure}
\includegraphics[width=0.8\columnwidth]{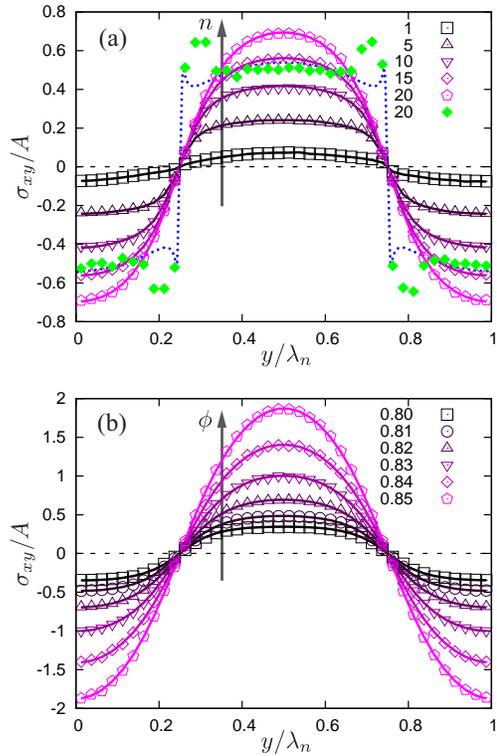}
\caption{(Color online)
The shear stress scaled by the amplitude, $A=10^{-3}d_0/t_0$,
where the open symbols are the results of MD simulations and the solid lines represent the non-local constitutive relations.
We increase (a) $n$ and (b) $\phi$ as listed in the legends and indicated by the arrows, where (a) $\phi=0.82$ and (b) $n=20$ are used.
The dotted line in (a) is the local constitutive relation, $\sigma_{xy}^\mathrm{L}(y)$, for $n=20$,
where the shear rate, $\dot{\gamma}(y)$, and area fraction, $\phi(y)$, are given by numerical solutions of the force balance equations (\ref{eq:fb_x}) and (\ref{eq:fb_y}).
The closed diamonds in (a) represent the local constitutive relation using $\dot{\gamma}(y)$ and $\phi(y)$ from MD simulations.
\label{fig:models}}
\end{figure}

To determine if particle migration accounts for wave-number dependence in the flow curves,
we assume $\sigma_{xy}^\mathrm{L}(y)=\eta_s[\phi(y),\dot{\gamma}(y)]\dot{\gamma}(y)$ and $\sigma_{yy}^\mathrm{L}(y)=\eta_c[\phi(y),\dot{\gamma}(y)]\dot{\gamma}(y)$ and numerically solve the force balance equations,
\begin{eqnarray}
\nabla_y\sigma_{xy}^\mathrm{L}(y) &=& -f_x^\mathrm{ex}(y)~, \label{eq:fb_x} \\
\nabla_y\sigma_{yy}^\mathrm{L}(y) &=& 0~, \label{eq:fb_y}
\end{eqnarray}
under boundary conditions, $\nabla_y\dot{\gamma}|_{y=0,L}=\nabla_y\phi|_{y=0,L}=0$,
where $f_x^\mathrm{ex}(y)=-(4\eta/\pi d_0^2)\phi(y)\left\{v_x(y)-A\sin q_ny\right\}$ represents the viscous force acting on the disks \cite{SupplMater}.
The dotted line in Fig.\ \ref{fig:models}(a) is the measured local shear stress, while the profiles $\dot{\gamma}(y)\equiv\nabla_y v_x(y)$ and $\phi(y)$ are given by the numerical solutions of Eqs.\ (\ref{eq:fb_x}) and (\ref{eq:fb_y}).
The local shear stress exhibits discontinuities around the shear-localized regions and significantly deviates from the results of MD simulation (the open pentagons in Fig.\ \ref{fig:models}(a)).
The discontinuities are due to the increase of $\phi$ above $\phi_J$ in the shear-localized regions (Fig.\ \ref{fig:profiles}(d)), generating a local yield stress.
We find that the local constitutive relation fails even if we take $\dot{\gamma}(y)$ and $\phi(y)$ from simulation data (the closed diamonds in Fig.\ \ref{fig:models}(a)).
We conclude that particle migration alone cannot account for the flow curves of Fig.~\ref{fig:setup}(b).
%

\emph{Non-local constitutive relations.}---
We now formulate non-local constitutive relations to describe shear localization and wave-number dependent flow behavior.
First, we introduce a general non-local constitutive relation as
\begin{equation}
\sigma_{xy}(y) = \int dy' \Theta(y,y')\dot{\gamma}(y')~,
\label{eq:sss}
\end{equation}
where $\Theta(y,y')$ represents non-local shear viscosity.
If the system is isotropic, the non-local shear viscosity can be normalized as $\Theta(y,y')\equiv\alpha(y-y')\eta_s\left[\phi(y'),\dot{\gamma}(y')\right]$,
where the propagator, $\alpha(l)$, is introduced as a symmetric function of the distance, $l\equiv y-y'$,
and is normalized as $\int_{-\infty}^\infty dl\alpha(l)=1$ \cite{Eringen}.
The shear stress, Eq.\ (\ref{eq:sss}), is then given by a weighted integral of the local shear stress,\
i.e.\ $\sigma_{xy}(y)=\int dy'\alpha(y-y')\sigma_{xy}^\mathrm{L}(y')=\int dl\alpha(l)\sigma_{xy}^\mathrm{L}(y-l)$.
Because the local constitutive relation is recovered if the propagator is replaced with Dirac's delta function,\ i.e.\ $\sigma_{xy}(y)=\sigma_{xy}^\mathrm{L}(y)$ if $\alpha(l)=\delta(l)$,
non-local effects can be quantified by a finite width of the propagator.

\begin{figure}
\includegraphics[width=\columnwidth]{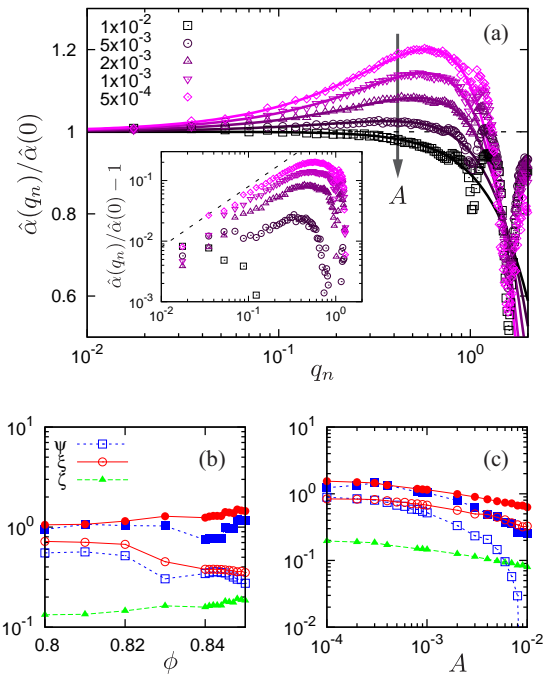}
\caption{(Color online)
(a) Semi-logarithmic plot of the propagator, $\hat{\alpha}(q_n)$, for varying forcing amplitude, $A$ (see legend).
The open symbols result from MD simulations, while the solid lines represent Eq.\ (\ref{eq:hat_alpha}).
The inset shows the double-logarithmic plot of $\hat{\alpha}(q_n)/\hat{\alpha}(0)-1$, where the dotted line has the slope $1$.
(b) and (c): The dependence of the non-local length scales, $\psi$, $\xi$, and $\zeta$, on (b) the area fraction, $\phi$, and (c) the forcing amplitude, $A$.
The open symbols are obtained by fitting Eq.\ (\ref{eq:hat_alpha}), and the closed symbols are found by fitting the stress profiles with a non-local constitutive model.
\label{fig:sigq}}
\end{figure}

Taking the Fourier transform of the non-local constitutive relation,
we find that the propagator is given by $\hat{\alpha}(q)=\hat{\sigma}_{xy}(q)/\hat{\sigma}_{xy}^\mathrm{L}(q)$ with the wave number, $q$,
where $\hat{\sigma}_{xy}(q)$ and $\hat{\sigma}_{xy}^\mathrm{L}(q)$ are Fourier coefficients of the non-local and local shear stress, respectively.
Note that the propagator goes to one in the long wavelength limit,\ i.e.\ $\hat{\alpha}(q)\rightarrow 1$ if $q\rightarrow 0$,
such that the local constitutive relation describes the shear stress for homogeneous (simple shear) flows.
Figure \ref{fig:sigq}(a) displays semi-logarithmic plots of the propagator (symbols) as a function of the imposed wave number, $q_n$,
where the Fourier coefficients, $\hat{\sigma}_{xy}(q_n)$ and $\hat{\sigma}_{xy}^\mathrm{L}(q_n)$, are obtained from the results of MD simulations.
We see that if the flow amplitude, $A$, is small enough, the propagator exhibits a small peak
(reminiscent of the ``dip" in the excess compliance of non-local elasticity \cite{nonl-el0}) before sharply decreasing.
Moreover, double-logarithmic plots (Fig.\ \ref{fig:sigq}(a), inset) imply a linear increase of the propagator for slow flows (dotted line).
This result is surprising because the propagator must be symmetric in $q_n$, and so the presence of a linear term implies that $\hat \alpha$ is non-analytic at zero wave number.
For small wave numbers, the propagator can be expanded as
\begin{equation}
\hat{\alpha}(q_n) \simeq \hat{\alpha}(0) + \psi|q_n| - (\xi q_n)^2~,
\label{eq:hat_alpha}
\end{equation}
where $\psi$ and $\xi$ are introduced as length scales encoding non-locality.
Note that the linear term, $\psi|q_n|$, is necessary to capture the peak for slow flows.
The solid lines in Fig.\ \ref{fig:sigq}(a) plots the expansion,\ Eq.\ (\ref{eq:hat_alpha}),
where we establish good agreement with numerical data by adjusting $\hat{\alpha}(0)$, $\psi$, and $\xi$.
We confirm that $\hat{\alpha}(0)=1\pm0.1$ ($\simeq 1$) and estimate the length scales for varying $A$ and $\phi$
(see SM \cite{SupplMater} for the dependence of $\hat{\alpha}(q)$ on $\phi$).

In previous studies \cite{sus-nonl0,sus-nonl1,sus-nonl2,foam-nonl0,nonl1,nonl2,nonl4}, a cooperative length was introduced to represent the range of non-locality.
Because this length depends sensitively on the system's proximity to a jamming or yielding transition, it has been widely accepted that non-locality has links to critical phenomena.
For our systems below jamming, the range of non-local effects is quantified by the length scales, $\psi$ and $\xi$, which we have quantitatively estimated by fitting Eq.\ (\ref{eq:hat_alpha}) to numerical results.
As shown in Figs.\ \ref{fig:sigq}(b) and (c), for the range of forcing amplitudes and area fractions accessed here,
these length scales never exceed a few particle diameters. Hence we find no evidence for a diverging nonlocal length scale below jamming.
%

\emph{Stress profiles.}---
We now demonstrate that the stress profiles from MD simulations can be captured within a nonlocal framework.
Inverting Eq.~(\ref{eq:hat_alpha}) is complicated by the non-analytic term.
If we neglect the peak in $\hat \alpha$, or if the forcing amplitude is sufficiently large that the peak vanshes, $\psi\simeq 0$, then
the expansion,\ Eq.\ (\ref{eq:hat_alpha}), can be inverted to obtain,
$\left\{1+(\xi q)^2\right\}\hat{\sigma}_{xy}(q)\simeq\hat{\sigma}_{xy}^\mathrm{L}(q)$.
In real space, this non-local constitutive relation becomes
\begin{equation}
\left(1-\xi^2\nabla_y^2\right)\sigma_{xy}(y)\simeq\sigma_{xy}^\mathrm{L}(y)~.
\label{eq:sss_aprox}
\end{equation}
Eq.\ (\ref{eq:sss_aprox}) is the inhomogeneous Helmholtz equation, where $\sigma_{xy}^\mathrm{L}(y)$ plays a role of the ``source".
Its solution is
\begin{equation}
\sigma_{xy}(y) = \frac{1}{2\xi}\int e^{-\frac{|y-y'|}{\xi}}\sigma_{xy}^\mathrm{L}(y')dy'~.
\label{eq:sss_long}
\end{equation}
This is an approximate form of the non-local constitutive relation,\ Eq.\ (\ref{eq:sss}),
where the propagator, $\alpha(l)$, is replaced with the exponential Green function, $e^{-\frac{|l|}{\xi}}/2\xi$, and $\xi$ is defined as the width of the propagator.

The approximate non-local constitutive relation,\ Eq.\ (\ref{eq:sss_aprox}), is associated with diffusion-type fluidity models \cite{sus-nonl0}
and an elastic counterpart of Eq.\ (\ref{eq:sss_aprox}) was recently reported \cite{nonl-el0}.
We find that in order to describe stress profiles accurately, Eq.\ (\ref{eq:sss_aprox}) must be generalized to fourth order as
$\left\{1-\xi^2\nabla_y^2+\left(\xi^4-\zeta^4\right)\nabla_y^4\right\}\sigma_{xy}(y)\simeq\sigma_{xy}^\mathrm{L}(y)$,
where $\zeta^4\equiv\int(l^4/4!)\alpha(l)dl$ is the fourth moment of the propagator.
In the SM \cite{SupplMater}, we present the solution for the fourth order propagator, analogous to Eq.~(\ref{eq:sss_long}), along with an approximate method to incorporate the influence of the peak in $\hat \alpha$ at low $A$.
As seen in Fig.~\ref{fig:models}, the stress profiles are in excellent agreement with the non-local constitutive relation, regardless of the wave number and area fraction.
Reassuringly, the $\phi$- and $A$-dependence of the non-local fitting parameters (Fig.~\ref{fig:sigq}(b) and (c), filled symbols) are compatible with the results of fitting Eq.~(\ref{eq:hat_alpha}) to the propagator.

\emph{Summary.}---
We have studied non-local effects in inhomogeneous Kolmogorov flows of soft athermal disks.
The rheology is strongly affected by the period of sinusoidal flow fields, and local constitutive relations fail even if particle migration is considered.
By introducing a general non-local constitutive relation, we quantitatively estimated the range of non-locality from the propagator.
Solutions for the stress profiles are in good agreement with simulations, provided the non-local constitutive relation is generalized to fourth order -- typical diffusion-type models fail to capture profiles for higher wave numbers.
Since most models for non-local effects \cite{sus-nonl0,sus-nonl1,sus-nonl2,foam-nonl0,nonl1,nonl2,nonl4} or shear-bands \cite{sband0,sband1} are diffusion-type,
our approach is an important step towards non-local continuum modeling of disordered materials \cite{Eringen}.
We find no evidence for critical divergence of the range of non-locality as jamming is approached from below -- non-local length scales remain on the order of the particle diameter for all sampled area fractions and flow rates.
We also note that the range of non-local {\em elastic} effects does not diverge under shear (though it does under compression) \cite{nonl-el0}.
As all studies reporting a diverging cooperativity length treated yield stress fluids \cite{nonlocal1,nonl1,nonl2,nonl4}, our results suggest that such divergence is associated with proximity to yielding, rather than jamming.

We thank K. Baumgarten for fruitful discussions.
This work was supported by KAKENHI Grant No.\ 16H04025 and No.\ 18K13464 from JSPS.
Some computations were performed at the Yukawa Institute Computer Facility, Kyoto, Japan.
BPT acknowledges support from the Dutch Organization for Scientific Research (NWO).
%
\bibliography{viscoelastic}
\end{document}